\documentclass{article}
\pdfoutput=1
\usepackage{url}
\usepackage{paralist}
\usepackage{color}
\usepackage{comment}
\usepackage{graphicx}
\usepackage{subcaption}

\newcommand{\C}{\mathcal{C}}

 \newcommand{\R}{\mathcal{R}}

\newcommand{\dd}[2]{#1_1,\ldots,#1_{#2}}             

  {$\Box$}

\newcommand{\rel}[1]{\mathsf{#1}}

\newcommand{\ext}[2]{#1^{#2}}

\newcommand{\dom}{\Gamma} 
\newcommand{\freshdom}{\Gamma_f}

\newcommand{\dep}{\Sigma}
\newcommand{\idep}{\Sigma_I}

\newcommand{\kdep}{\Sigma_K}

\newcommand{\key}[1]{\mathit{key}(#1)}
\newcommand{\isa}[1]{\mathit{ISA}}

\newcommand{\arity}[1]{\mathit{arity}(#1)}

\newcommand{\ins}[1]{\bar{#1}}

\newcommand{\insX}{\ins{X}}
\newcommand{\insY}{\ins{Y}}

\newcommand{\chase}[2]{\mathit{chase}_{#1}(#2)}

\newcommand{\level}[1]{\mathit{level}(#1)}

\newcommand{\maxlevel}[0]{\delta_{M}}

\newcounter{cefalo}

\newcounter{cefalocont}

\newtheorem{theorem}{Theorem}

\newtheorem{definitionAux}[theorem]{Definition} 
\newenvironment{definition}{\begin{definitionAux} 
}{\end{definitionAux}}

\newtheorem{exampleAux}{Example}
\newenvironment{example}{\begin{exampleAux}\upshape}
  {\markfull\end{exampleAux}}

\def\qed{\hfill{\qedboxempty}      
  \ifdim\lastskip<\medskipamount \removelastskip\penalty55\medskip\fi}

\def\qedboxempty{\vbox{\hrule\hbox{\vrule\kern3pt
                 \vbox{\kern3pt\kern3pt}\kern3pt\vrule}\hrule}}

\def\qedfull{\hfill{\qedboxfull}   
  \ifdim\lastskip<\medskipamount \removelastskip\penalty55\medskip\fi}

\def\qedboxfull{\vrule height 4pt width 4pt depth 0pt}

\newcommand{\markfull}{\qedboxfull}
\newcommand{\markempty}{\qed}

\newcommand{{\incolumn}}[1]{\begin{tabular}[c]{c} #1 \end{tabular}}
\newcommand{{\incolumnmath}}[1]{\begin{array}[c]{c} #1 \end{array}}

\usepackage{amssymb}
\begin{document}

\title{On the dependency on the size of the data\\
when chasing under conceptual dependencies}
\author{Davide Martinenghi\\\\
Politecnico di Milano\\
Piazza Leonardo da Vinci 32\\
20133 Milano, Italy\\
\url{davide.martinenghi@polimi.it}}
\date{}

\maketitle

\begin{abstract}
	Conceptual dependencies (CDs) are particular kinds of key dependencies (KDs) and inclusion dependencies (IDs) that precisely characterize relational schemata modeled according to the main features of the Entity-Relationship (ER) model.
	An instance for such a schema may be inconsistent (data violate the dependencies) and incomplete (data constitute a piece of correct information, but not necessarily all the relevant information).
	While undecidable under general KDs and IDs, query answering under incomplete data is known to be decidable for CDs.
	The known techniques are based on the chase -- a special instance, organized in levels of depth, that is a representative of all the instances that satisfy the dependencies and that include the initial instance. 
	Although the chase generally has infinite size, query answering can be addressed by posing the query (or a rewriting thereof) on a finite, initial part of the chase.
	Contrary to previous claims, we show that the maximum level of such an initial part cannot be bounded by a constant that does not depend on the size of the initial instance.
\end{abstract}
%
%
%

\section{Introduction}
\label{sec:intro}

In the context of conceptual data models, particularly the Entity-Relationship (ER) model~\cite{Chen76} and its variants, data may be inconsistent and incomplete with respect to the constraints imposed on the model.

Among the many variants of the ER model, a conceptual model of interest was presented in~\cite{Cali06} with the ability to represent classes of objects with their attributes, relationships among classes, cardinality constraints in the participation of entities in relationships, and is-a relations among both classes and relationships.
Such a model, called Extended ER (EER) model, can be formalized by means of constraints called conceptual dependencies (CDs).

In the presence of incomplete data with respect to the CDs associated with the EER schema, according to the so-called \emph{sound semantics} (see, e.g., \cite{CaLR03}), one only considers databases that are supersets of the initial data, and satisfy the constraints.
Given a query, the \emph{certain answers} are those that are true in all such databases.

The problem of query answering under CDs in the presence of incomplete information under the sound semantics has been addressed in~\cite{CM:TPLP2010},
where the initial query is rewritten into a new query that takes into account the constraints, and such that its evaluation over the initial incomplete data returns the certain answers.
The rewriting is heavily based on the chase, which is a formal tool for query answering with \emph{incomplete data}.
The result of the chase is a new database, also called chase.
More specifically, the chase is a (potentially infinite) database, organized in levels, whose construction amounts to repairing violations of IDs and KDs, the former by adding tuples, and the latter by merging tuples.
The presented technique may have practical interest if the rewriting operates at a purely \emph{intensional level}, reasoning on queries and constraints, and only querying the data at the last step, since the size of the data is usually much larger than the size of the constraints.
However, the rewriting is based on an encoding of the first few levels of the chase.
We show that, in general, such number of levels cannot be bounded by a constant that does not depend on the size of the initial data.
This amends and supersedes an opposite statement made in~\cite{Cali06}.

\section{Preliminaries}
\label{sec:prelim}

%
We refer to~\cite{Lloy87,AbHV95,CM:TPLP2010,Reit78,lenz02} for common notions about relational databases, such as relational schemata, conjunctive queries, homomorphisms, integrity constraints and satisfaction thereof.
In the following we often refer to a domain of constants $\dom$ along with a domain of fresh constants $\freshdom$.

\subsection{Dependencies}

In this paper we consider the following kinds of integrity constraints:
\begin{compactenum}[\itshape (i)]
	\item \textit{Inclusion dependencies (IDs).}
	An inclusion dependency $\sigma_I$ between relational predicates $r_1$ and $r_2$ is denoted by $r_1[\ins{X}] \subseteq r_2[\ins{Y}]$.
	Given a database $D$ with values only in $\dom$, such a constraint is satisfied in $D$, written $D\models\sigma_I$, iff, for each tuple $t_1$ in $\ext{r_1}{D}$, there exists a tuple $t_2$ in $\ext{r_2}{D}$ such that $t_1[\ins{X}]=t_2[\ins{Y}]$.
	An ID is said to be a \emph{full-width} ID if every attribute of $r_1$ occurs in $\ins{X}$ exactly once and every attribute of $r_2$ occurs in $\ins{Y}$ exactly once.
	\item \textit{Key dependencies (KDs).}
	A key dependency $\sigma_K$ over a relational predicate $r$ with $\arity{r}\geq 2$ is denoted by $\key{r}=\ins{K}$, where $\ins{K}$ is a nonempty subset of the attributes of $r$.
	Given a database $D$ with values only in $\dom$, such a constraint is satisfied in $D$, written $D\models\sigma_K$, iff, for each $t_1,t_2\in\ext{r}{D}$ such that $t_1\neq t_2$, we have $t_1[\ins{K}^*]\neq t_2[\ins{K}^*]$, where $\ins{K}^*$ is any sequence of $|\ins{K}|$ attributes where each attribute in $\ins{K}$ occurs exactly once.
\end{compactenum}

We restrict our attention to the so-called \emph{certain answers} to a query: given a finite database $D$, the answers we consider are those that are true in all models, i.e., in \emph{all} the databases that contain $D$ \emph{and} satisfy the dependencies.
In the following, we shall always assume that the initial database has finite size, while no finiteness assumptions is made on the models.
%
%
%


\subsection{The Conceptual Model}
\label{sec:conceptual-model}

The conceptual model we adopt in this paper is called \emph{Extended Entity-Relationship (EER) model}~\cite{CM:TPLP2010}.
Such a model is an extension of the one presented in~\cite{CCDL01e} and incorporates the basic features of the ER model~\cite{Chen76} and OO~models, including subset (or is-a) constraints on both entities and relationships.

\begin{figure}[tb]
  \centering \includegraphics{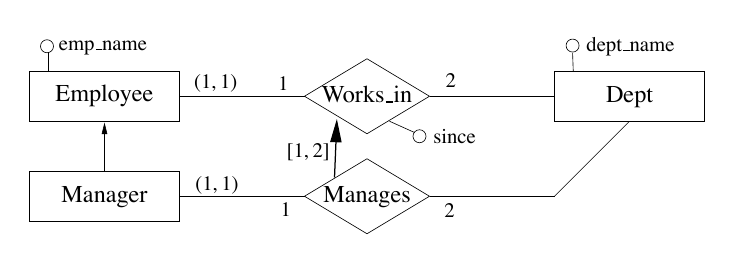}
  \caption{An EER schema}
  \label{fig:ER}
\end{figure}

Like an ER schema, an \emph{EER schema} consists of a collection of entity, relationship, and attribute definitions.
An example EER schema is shown in Figure~\ref{fig:ER}.

Instead of characterizing EER schemata with an ad hoc language, we represent them directly in relational terms, as shown in the following example taken from~\cite{CM:TPLP2010}.
\begin{example}\label{exa:ER-query-contd} Consider the EER schema shown in
  Figure~\ref{fig:ER}. The set of IDs and KDs that completely characterize such a schema consists of the following dependencies:
	$$\begin{array}{rrcll}
		\sigma_1:&\rel{dept\_name}[1]&\subseteq& \rel{dept}[1] \\
		\sigma_2:&\rel{emp\_name}[1]&\subseteq& \rel{employee}[1] \\
		\sigma_3:&\rel{since}[1,2]&\subseteq& \rel{works\_in}[1,2] \\
		\sigma_4:&\rel{works\_in}[1]&\subseteq& \rel{employee}[1] \\
		\sigma_5:&\rel{works\_in}[2]&\subseteq& \rel{dept}[1] \\
		\sigma_6:&\rel{manages}[1]&\subseteq& \rel{manager}[1] \\
		\sigma_7:&\rel{manages}[2]&\subseteq& \rel{dept}[1] \\
		\sigma_8:&\rel{manager}[1]&\subseteq& \rel{employee}[1] \\
		\sigma_9:&\rel{manages}[1,2]&\subseteq& \rel{works\_in}[1,2] \\
		\sigma_{10}:&\rel{employee}[1]&\subseteq& \rel{works\_in}[1] \\
		\sigma_{11}:&\rel{manager}[1]&\subseteq& \rel{manages}[1] \\
		\sigma_{12}:&\key{\rel{works\_in}}&=&\{1\} \\
		\sigma_{13}:&\key{\rel{manages}}&=&\{1\}
  \end{array}$$
\end{example}

The IDs and KDs that fully encode EER schemata are precisely what we refer to as conceptual dependencies, as captured by the following (rather technical) definition (stated as a Proposition in~\cite{CM:TPLP2010})
\begin{definition}\label{pro:CDs}
	Consider a schema $\R$ and a set of dependencies $\dep=\idep \cup \kdep$, where $\idep$ is a set of inclusion dependencies and $\kdep$ is a set of key dependencies expressed over $\R$.
	Then, $\Sigma$ is a set of \emph{Conceptual Dependencies} (CDs) if and only if we can partition $\R$ in three sets $\R_R$, $\R_E$, and $\R_A$ such that the following holds.
	\begin{compactenum}[\itshape (a)]
		\item All predicate symbols in $\R_E$ are unary.
		\item All predicate symbols in $\R_R$ and $\R_A$ have arity at least 2.
		\item The dependencies in $\kdep$ have one of the following forms
		\begin{compactenum}[\itshape (1)]
			\item $\key{r}=\{i\}$, with $1\leq i\leq \arity{r}$, where $r\in\R_R$.
			\item $\key{a}=\{1,\ldots,n\}$, where $a\in\R_A$ and $n=\arity{a}-1$.
		\end{compactenum}
		\item The dependencies in $\idep$ have one of the following forms
		\begin{compactenum}[\itshape (1)]
			\item $e_1[1]\subseteq e_2[1]$, where $\{e_1,e_2\}\subseteq\R_E$.
			\item $e[1]\subseteq r[i]$, where $e\in\R_E$, $r\in\R_R$, and $1\leq i\leq \arity{r}$.
			\item $r[i]\subseteq e[1]$, where $r\in\R_R$, $e\in\R_E$, and $1\leq i\leq \arity{r}$.
			\item $r_1[1,\ldots,k]\subseteq r_2[i_1,\ldots,i_k]$, where $\{r_1,r_2\}\subseteq\R_R$, $\arity{r_1}=\arity{r_2} = k$, and $(i_1,\ldots,i_k)$ is a permutation of $(1,\ldots,k)$.
			\item $a[1]\subseteq e[1]$, where $a\in\R_A$ and $e\in\R_E$.
			\item $a[1,\ldots,n]\subseteq r[1,\ldots,n]$, where $a\in\R_A$, $r\in\R_R$, and $n=\arity{r}=\arity{a}-1$.
			\item $e[1]\subseteq a[1]$, where $e\in\R_E$ and $a\in\R_A$.
			\item $r[1,\ldots,n]\subseteq a[1,\ldots,n]$, where $r\in\R_R$, $a\in\R_A$, and $n=\arity{r}=\arity{a}-1$.
		\end{compactenum}
		\item For every predicate $r\in\R_R$ and for $1\leq i\leq \arity{r}$, there exists an ID $r[i]\subseteq e_i[1]$ in $\idep$ such that $e_i\in\R_E$ and there is no $e'_i\in\R_E$, with $e_i\neq e'_i$, such that $r[i]\subseteq e'_i[1]$ is in $\idep$.
		\item For every predicate $a\in\R_A$, there exists an ID $a[1,\ldots,n]\subseteq p[1,\ldots,n]$ in $\idep$ such that $p\in\R_R\cup \R_E$ and $n=\arity{p}=\arity{a}-1$, and there is no $p'\in\R_R\cup\R_E$, with $p\neq p'$, such that $a[1,\ldots,n]\subseteq p'[1,\ldots,n]$ is in $\idep$.
		\item For every ID $e[1]\subseteq r[i]$ in $\idep$, with $e\in\R_E$, $r\in\R_R$, and $1\leq i\leq \arity{r}$, there is an ID $r[i]\subseteq e[1]$ in $\idep$.
		\item For every ID $r[1,\ldots,n]\subseteq a[1,\ldots,n]$ in $\idep$, with $r\in\R_R$, $a\in\R_A$, and $n=\arity{r}=\arity{a}-1$, there is an ID $a[1,\ldots,n]\subseteq r[1,\ldots,n]$ in $\idep$.
		\item For every ID $e[1]\subseteq a[1]$ in $\idep$, with $e\in\R_E$, $a\in\R_A$, and $\arity{a}=2$, there is an ID $a[1]\subseteq e[1]$ in $\idep$.
	\end{compactenum}
\end{definition}
The problem of querying incomplete databases under KDs and IDs is in general undecidable~\cite{Cali03t,CaLR03}.
The largest subclass of functional dependencies\footnote{Functional dependencies are a generalization of key dependencies~\cite{AbHV95}.} and IDs for which query answering is known to be decidable is the class of keys and non-key conflicting IDs~\cite{Cali03t,CaLR03}.
In~\cite{CM:TPLP2010}, a technique for solving the problem of querying incomplete databases under CDs is shown. Such a technique is based on the notion of chase, and consists in rewriting the given query so that the evaluation of the rewritten query returns the certain answers.


\subsection{Chase}
\label{sec:chase}

In this section we introduce the notion of \emph{chase}, which is a fundamental tool for dealing with database constraints~\cite{MaMS79,MaSY81,Vard83,JoKl84}.

Intuitively, given a database, its facts in general do not satisfy the dependencies.
The idea of the chase is to convert the initial facts into a new set of facts constituting a database that satisfies the dependencies, possibly by collapsing facts (according to KDs) or adding new facts (according to IDs).
When new facts are added, some of the constants need to be \emph{fresh}, as we shall see in the following.
Next follows an adaptation of the well-known chase rules for functional dependencies and IDs~\cite{JoKl84} to the simpler case of KDs and IDs, and some results about query answering over the chase that were presented in~\cite{CM:TPLP2010}.

Let $D$ be the set of facts before the application of a rule.

\textsc{Inclusion Dependency Chase Rule.}
Let $r,s$ be relational symbols in $\R$.
Suppose there is a tuple $t$ in
$r^{D}$, and there is an ID $\sigma \in \idep$ of the form $r[\ins{X}_r] \subseteq s[\ins{X}_s]$.
If there is no tuple $t'$ in $s^{D}$ such that $t'[\insX_s]=t[\insX_r]$ (in this case we say the rule is \emph{applicable}), then we add a new tuple $t_{\mathit{chase}}$ in $s^{D}$ such that $t_{\mathit{chase}}[\insX_s]=t[\insX_r]$, and for every attribute $A_i$ of $s$ such that
$A_i \notin \insX_s$, $t_{\mathit{chase}}[A_i]$ is a fresh value in $\freshdom$ that \emph{follows}, according to lexicographic order, all the values already present in the chase.
Note also that we assume that all the values in $\freshdom$ follow, according to lexicographic order, all the values in $\dom$.

\textsc{Key Dependency Chase Rule.}
Let $r$ be a relational symbol in $\R$.
Suppose there is a KD $\kappa$ of the form $\key{r} = \insX$.
If there are two \emph{distinct} tuples
$t,t' \in r^{D}$ such that $t[\insX] = t'[\insX]$ (in this case we say the rule is \emph{applicable}), make the symbols in $t$ and $t'$ equal in the following way.
Let $\insY=\dd{Y}{\ell}$ be the attributes of $r$ that are not in $\insX$; for all $i\in\{1,\ldots,\ell\}$, make $t[Y_i]$ and $t'[Y_i]$ merge into a combined symbol according to the following criterion:
\begin{inparaenum}[\itshape (i)]
	\item if both are constants in $\dom$ and they are not equal, the rule fails to apply and the chase construction process is halted; 
	\item if one is in $\dom$ and the other is a fresh constant in $\freshdom$, let the combined symbol be the non-fresh constant;  
	\item if both are in $\freshdom$, let the combined symbol be the one preceding the other in lexicographic order.
\end{inparaenum}
Finally, replace all occurrences in
$D$ of $t[Y_i]$ and $t'[Y_i]$ with their combined symbol.

Now we come to the formal definition of the chase, which uses the notion of \emph{level} of a tuple; intuitively, the lower the level of a tuple, the earlier the tuple has been constructed in the chase.
In order to make all steps in the construction of the chase univocally determined by the definition, we assume that all facts can be sorted according to lexicographic order (e.g., by using a string comprising the predicate name and the names of all constants in the fact), and so can all pairs of facts as well as all dependencies (e.g., also by using strings that encode them).
%
\begin{definition}[Chase] \label{def:chase}
	Let $D$ be a database for a schema $\R$, and $\dep$ a set of CDs.
	We call \emph{chase} of $D$ according to $\dep$, denoted $\chase{\dep}{D}$, the database constructed from $D$ by  repeatedly executing the following steps, while the KD and ID chase rules are applicable; every tuple $t\in\chase{\dep}{D}$ is also assigned a \emph{level}, denoted by $\level{t}$; if $t\in D$, then $\level{t}=0$.
	\begin{asparaenum}[\itshape (1)]
		\item While there are pairs of facts on which the KD chase rule is applicable, 
		take the pair $t_1,t_2$ such that $min(\level{t_1},\level{t_2})$ is minimal (if there is more than one, take the pair that comes first in lexicographic order) and apply the KD chase rule on $t_1,t_2$ w.r.t. a KD $\kappa$ (if there is more than one KD for which the KD chase rule is applicable on $t_1,t_2$, take the KD that comes first in lexicographic order) so that $t_1,t_2$ collapse into a fact $t_3$; if the rule fails, the chase cannot be constructed and, thus, does not exist; else we define $\level{t_3}=min(\level{t_1},\level{t_2})$.
		\item If there are facts on which the ID chase rule is applicable w.r.t. a full-width ID, choose \emph{the one} (say $t'$) at the lowest level that lexicographically comes first and apply the ID chase rule on $t'$ w.r.t. a full-width ID $\sigma$ (if there is more than one full-width ID for which the ID chase rule is applicable on $t'$, take the full-width ID that comes first in lexicographic order) to generate a new fact $t''$; else, if there are facts on which the ID chase rule is applicable, choose \emph{the one} (say $t'$) at the lowest level that lexicographically comes first and apply the ID chase rule on $t'$ w.r.t. an ID $\sigma$ (if there is more than one ID for which the ID chase rule is applicable on $t'$, take the ID that comes first in lexicographic order) to generate a new fact $t''$.
		We define $\level{t''}=\level{t'}+1$.
	\end{asparaenum}
\end{definition}

Note that, according to Definition~\ref{def:chase}, the chase is constructed by applying the KD chase rule as long as possible, then the ID chase rule exactly once, then the KD chase rule as long as possible, etc., until no more rule is applicable.

As we pointed out before, the aim of the construction of the chase is to make the initial database satisfy the KDs and the IDs, by repairing the violations of the constraints.
The obtained (possibly infinite) instance is a representative of all databases that are a superset of the initial database
and satisfy the constraints.
Notice that key dependency violations cannot be repaired by constructing a chase, but would require an explicit treatment; in such a case the chase does not exist.
It is easy to see that $\chase{\dep}{D}$ can be infinite only if the set of IDs in $\dep$ is \emph{cyclic}~\cite{AbHV95,JoKl84}, i.e., if there is a sequence of IDs in $\dep$ of the form $r_1[\insX_1]\subseteq r_2[\insX_1'], r_2[\insX_2]\subseteq r_3[\insX_2'], \ldots, r_n[\insX_n]\subseteq r_{n+1}[\insX_n']$ and $r_{n+1}=r_1$.
An example of chase is shown next.
\begin{example} \label{exa:chase} Consider the dependencies of
  Example~\ref{exa:ER-query-contd}.  Suppose we have an initial (incomplete)
  database, with the facts $\rel{manager}(m)$ and $\rel{works\_in}(m,d)$.  If
  we construct the chase, we obtain the facts $\rel{employee}(m)$,
  $\rel{manages}(m,\alpha_1)$, $\rel{works\_in}(m,\alpha_1)$,
  $\rel{dept}(\alpha_1)$, where $\alpha_1$ is a fresh constant.  Observe that
  $m$ cannot participate more than once in $\rel{works\_in}$, so we deduce
  $\alpha_1 = d$.  We must therefore replace $\alpha_1$ with $d$ in the rest
  of the chase, including the part that has been constructed so far.
Therefore, $\chase{\dep}{D}=\{\rel{manager}(m),$ $ \rel{works\_in}(m,d),$ $ \rel{employee}(m),$ $ \rel{manages}(m,d), \rel{dept}(d)\}$.
\end{example}

\section{Dependency on the size of the data}
\label{sec:size}

In~\cite{JoKl84}, a well-known technique was presented for checking the containment relationship $Q_1 \subseteq_{\dep} Q_2$ between two conjunctive queries $Q_1$ and $Q_2$ under a set $\dep$ of functional and inclusion dependencies.
Query $Q_1$ is ``frozen'', i.e., all its atoms are turned into facts by sending variables into fresh constants.
The chase $\C$ of the frozen body of $Q_1$ is a representative of all databases $B$ that answer the query $Q_1$ and that satisfy the constraints, in the sense that, for every such $B$, there is a homomorphism $\lambda$ from $\C$ to $B$.
Containment holds if and only if there is a query homomorphism sending the body of $Q_2$ into $\C$ and the head of $Q_2$ into the corresponding frozen head of $Q_1$.

In~\cite{Cali06}, a technique is described for checking conjunctive query containment under CDs.
Although the chase may have infinite size, under CDs only a finite portion of the chase (up to a certain level), is relevant for query answering as well as containment checking purposes.
In particular, if there is a homomorphism $\mu$ sending the body of a conjunctive query $Q$ into facts of the chase of $D$ and the head variables $\vec X$ into the answer tuple $\vec t$, then there is another homomorphism $\mu'$ sending the body of $Q$ into facts of the chase of $D$ \emph{at a level less than} $\maxlevel$ and, again, $\vec X$ into $\vec t$.
By the above results, under CDs, only the first $\maxlevel$ levels of the chase need to be considered to check containment. However, the construction of the first $\maxlevel$ levels might require to go deeper in the construction of the chase, since the application of the KD chase rule might propagate constants from greater to lower levels. This back-propagation can, however, only apply for at most $\maxlevel$ levels. All these considerations together entail decidability of conjunctive query containment under CDs.

A polynomial-time complexity with respect to the size of $Q_1$ is claimed in~\cite{Cali06}. This is also regarded as a \emph{data complexity}, since the frozen query plays the role of the initial database.
In particular, it is stated (Lemma 2 in~\cite{Cali06}) that the level $\maxlevel$ does not depend on the size of the initial database.
Furthermore, it is stated (Lemma 3 in~\cite{Cali06}) that constants occurring in the chase at a level $l$ do not occur anymore in the chase at levels greater than $l+\delta$, where $\delta$ is a constant that does not depend on the size of the initial database.

We show how chasing under CDs may propagate constants in the initial database to facts in the chase whose level depends on the size of the initial database. This contradicts both Lemma 2 and Lemma 3 in~\cite{Cali06}, since both $\maxlevel$ and $\delta$ must depend on the size of the initial database.

Consider the following set $\dep = \kdep \cup \idep$, where
$$\kdep = \{ \key{s} = \{1\} \}$$
and
$$\begin{array}{rll}\idep = \{
& r[1] \subseteq e[1], \\
& r[2] \subseteq e[1], \\
& r[1,2] \subseteq s[1,2],\\
& e[1] \subseteq r[1],\\
& s[2] \subseteq e'[1],\\
& s[1] \subseteq e'[1] & \},
\end{array}$$
and a database
$$D = \{e(1), s(1,2), s(2,3), \ldots, s(n-1,n)\}.$$
Note that $e'$ and the last three IDs in $\idep$ have no special role except to ensure that $\dep$ is a set of CDs.
In the chase of $D$ with respect to $\dep$, each constant $i$, $1\leq i\leq n$ occurs at least in a fact at level $2(i-1)$.
The construction of the chase is shown in Figure~\ref{fig:chase} as a forest-like structure, where the levels of the facts are indicated in the side margin in gray.

It is easily seen that $n$ applications of the ID chase rule are required to ``close off'' the initial facts of the form $s(k-1,k)$, with $2\leq k\leq n$, by generating $e'(1)$ as well as all facts of the form $e'(k)$ at level $1$.
Furthermore, $4$ applications are required to generate a fact of the form $e(k)$ from a fact of the form $e(k-1)$.
Figure~\ref{fig:chase0} shows a chase structure in which $e(i-1)$ has been generated at level $2i-4$.
The ID chase rule is applied to $e(i-1)$ (highlighted in blue) and generates a fact of the form $r(i-1,\alpha)$, where $\alpha$ is a fresh constant, at level $2i-3$.
Figure~\ref{fig:chase1} shows the subsequent application of the ID chase rule to $r(i-1,\alpha)$ (in blue), which generates a fact $s(i-1,\alpha)$ at level $2i-2$.
Then, the KD chase rule is applied, as shown in Figure~\ref{fig:chase2}, on the fact $s(i-1,i)$ at level $0$ and the newly generated $s(i-1,\alpha)$ (shown in red), which enforces the substitution of $\alpha$ with $i$ and the elimination of $s(i-1,\alpha)$ from the chase.
The resulting structure is shown in Figure~\ref{fig:chase3}. Finally, the ID chase rule is applied on $r(i-1,i)$ (in blue) and generates $e(i)$ at level $2i-2$, as shown in Figure~\ref{fig:chase4}. The construction will continue with an application of the ID chase rule on $e(i)$.
The same argument can be applied until a fact $e(n)$ is generated at level $2n-2$.
At this point, the construction continues by generating facts at levels greater than $2n-2$ but will never affect any of the lower levels, since:
\begin{inparaenum}[\itshape i)]
	\item no ID chase rule is applicable on facts at a level less than $2n-2$,
	\item no KD chase rule is applicable on any two facts at a level less than $2n-2$, and
	\item if a KD chase rule between one fact at level less than $2n-2$ and one at level greater than $2n-2$ is applicable, the latter is eliminated, while the former is kept, and no fact in the first $2n-2$ levels is affected, since no fresh constant occurs in them.
\end{inparaenum}

Since the number $n$ was chosen arbitrarily, $D$ contains exactly $n$ facts, and the database constant $n$ occurs at level $2n-2$ in the chase, we can conclude that neither $\maxlevel$ nor $\delta$ can be chosen independently of the size of the database. We show this by contradiction.

Suppose that $\maxlevel$ is independent of the size of the database. Then, $n$ can be chosen in the previous example so that $2n-2>\maxlevel$. Since $e(n)$ does not occur at a level less than $2n-2$, the answer to a query of the form $q(X)\leftarrow e(X)$ posed over the first $\maxlevel$ levels of the chase does not contain the tuple $\langle n \rangle$.
If the same query is posed over the entire chase, the answer does contain $\langle n \rangle$. Therefore the facts occurring at levels greater than $\maxlevel$ are relevant for query answering. Contradiction.

Suppose now that $\delta$ is independent of the size of the database. Again, $n$ can be chosen in the previous example so that $2n-2>\delta$.
However, there is a constant ($n$) that occurs both at level $0$ (in $s(n-1,n)$) and at level $2n-2>0+\delta$ (in $e(n)$). Contradiction.

\newcommand{\scalefactor}{1} 

\begin{figure}
    \begin{subfigure}{\textwidth}
        \includegraphics[width=\scalefactor\columnwidth]{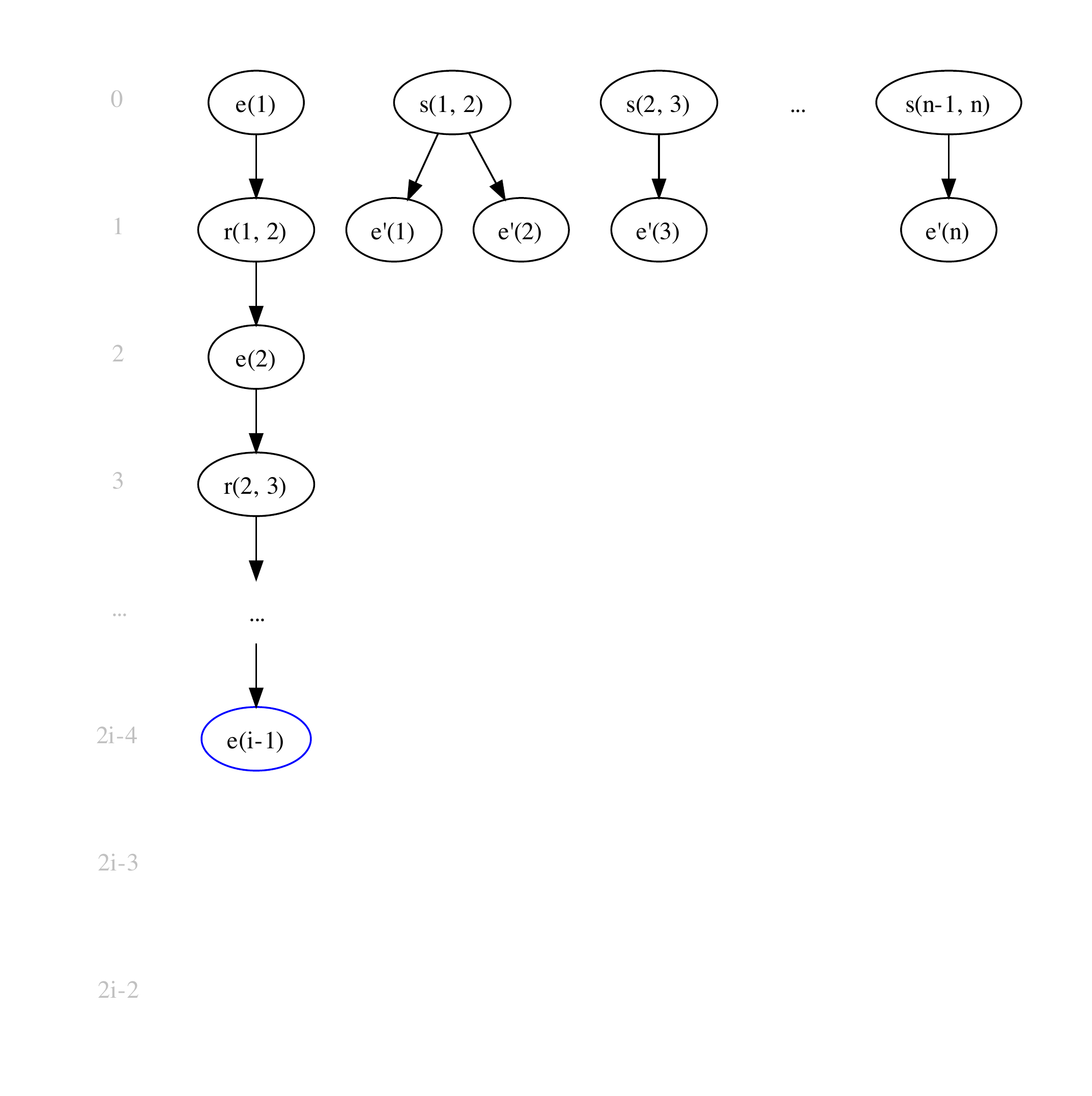}
        \caption{$e(i-1)$ has been generated at level $2i-4$. The ID chase rule will be applied to $e(i-1)$ (highlighted in blue)}\label{fig:chase0}
    \end{subfigure}
\caption{Steps of the construction of the chase. Facts selected for
  the application of a chase rule have a different color (blue: ID
  chase rule, red: KD chase rule).} \label{fig:chase}
\end{figure}
\clearpage
\begin{figure}
    \ContinuedFloat 
    \begin{subfigure}{\textwidth}
        \includegraphics[width=\scalefactor\columnwidth]{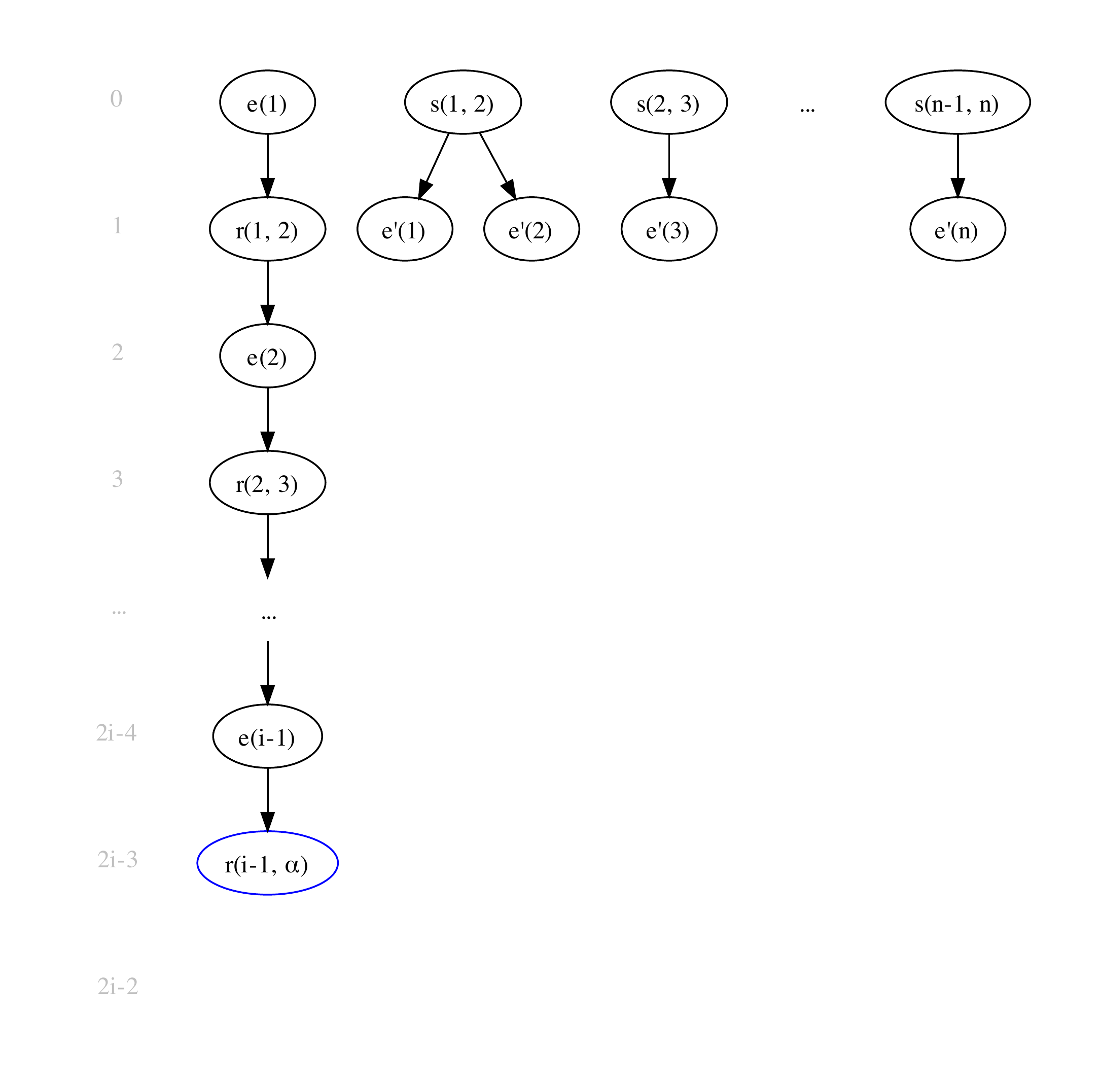}
        \caption{$r(i-1,\alpha)$ is generated at level $2i-3$. The ID chase rule will be applied to $r(i-1,\alpha)$ (in blue)}\label{fig:chase1}
    \end{subfigure}
\end{figure}
\clearpage
\begin{figure}
    \ContinuedFloat 
    \begin{subfigure}{\textwidth}
        \includegraphics[width=\scalefactor\columnwidth]{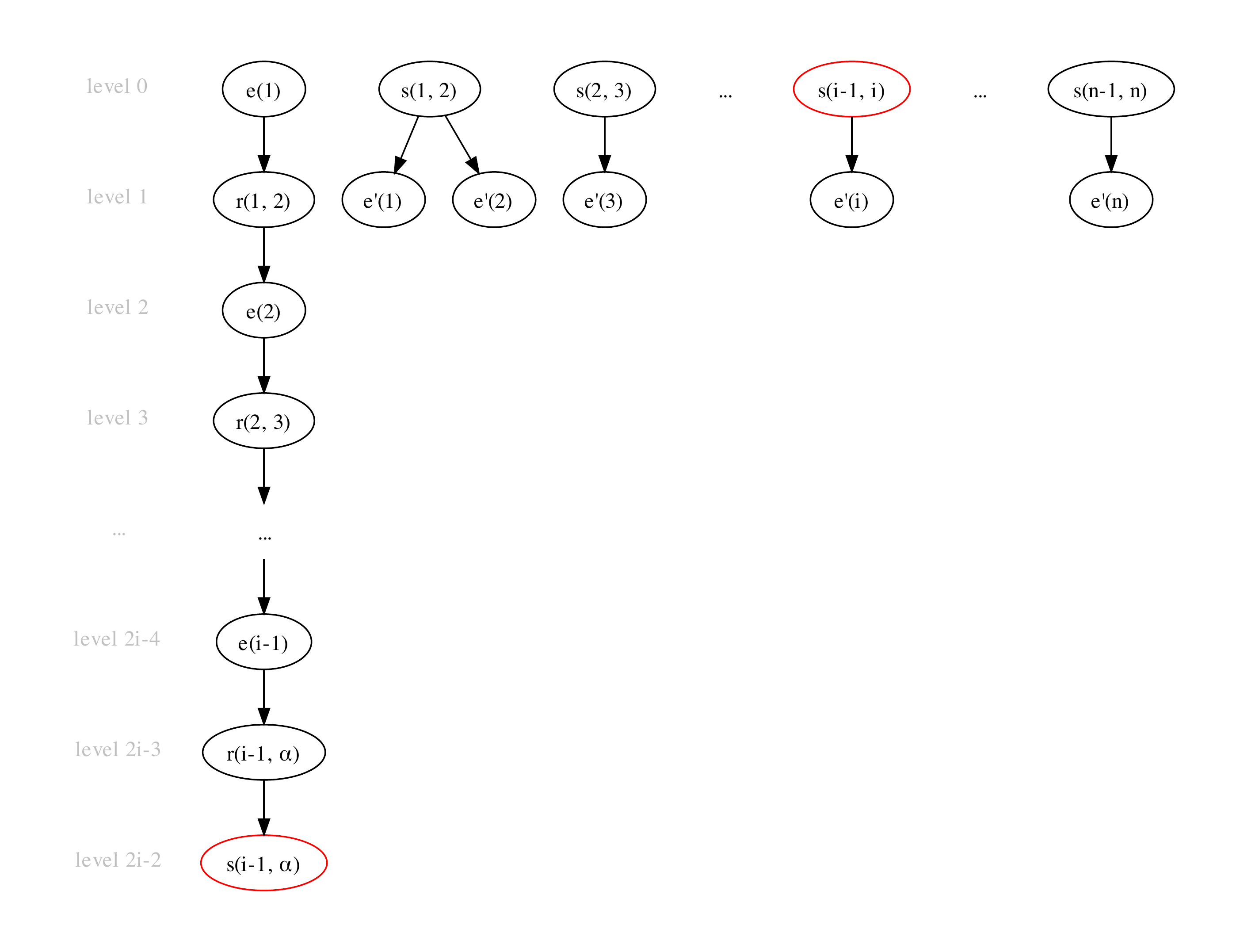}
       \caption{$s(i-1,\alpha)$ is generated at level $2i-2$. The KD chase rule will be applied to $s(i-1,i)$ at level $0$ and $s(i-1,\alpha)$ (shown in red)}\label{fig:chase2}
    \end{subfigure}
\end{figure}
\clearpage
\begin{figure}
    \ContinuedFloat 
    \begin{subfigure}{\textwidth}
        \includegraphics[width=\scalefactor\columnwidth]{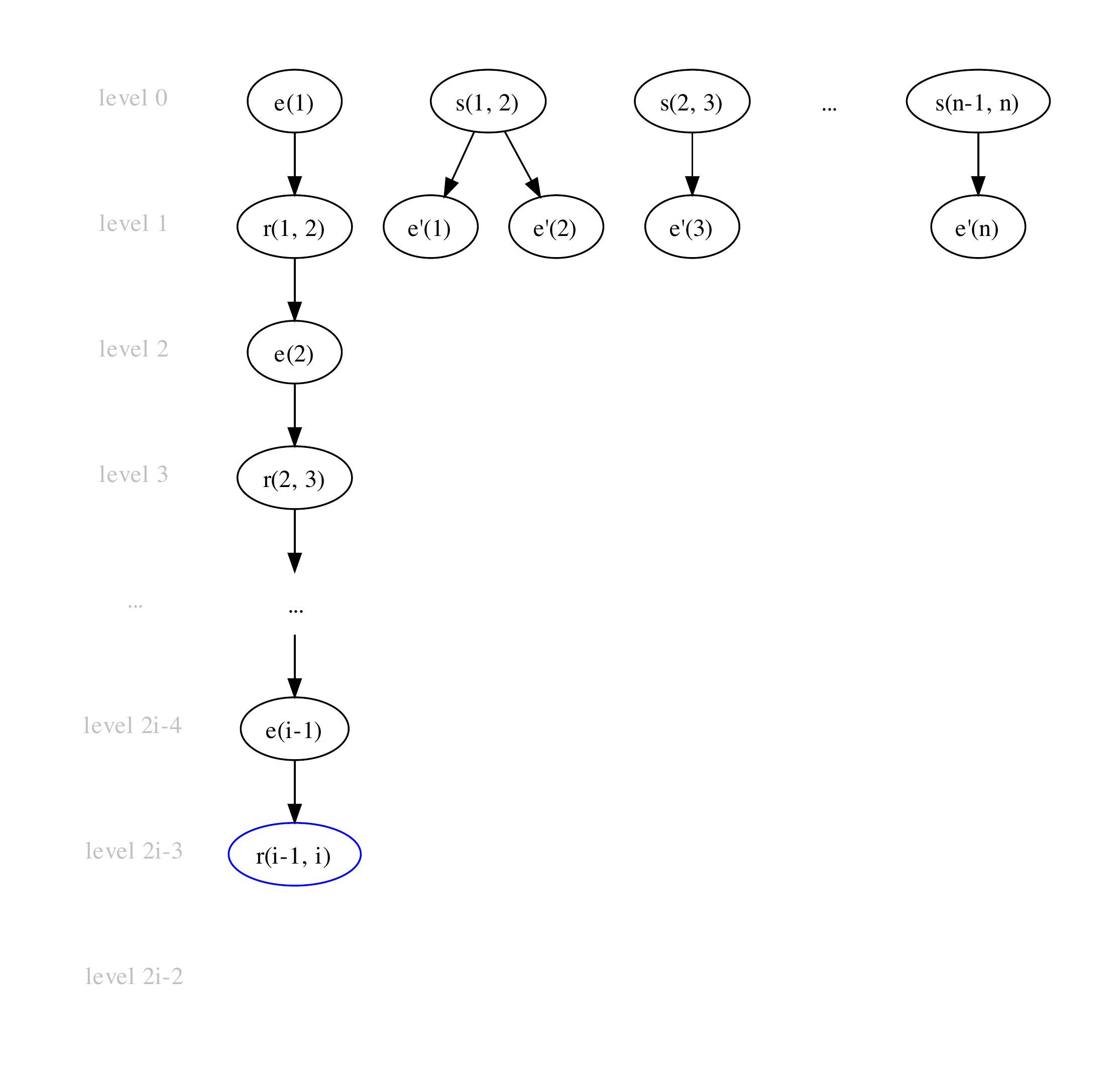}
        \caption{$\alpha$ is replaced by $i$ and $s(i-1,\alpha)$ is eliminated. The ID chase rule is applied on $r(i-1,i)$ (in blue)}\label{fig:chase3}
    \end{subfigure}
\end{figure}
\clearpage
\begin{figure}
    \ContinuedFloat 
    \begin{subfigure}{\textwidth}
        \includegraphics[width=\scalefactor\columnwidth]{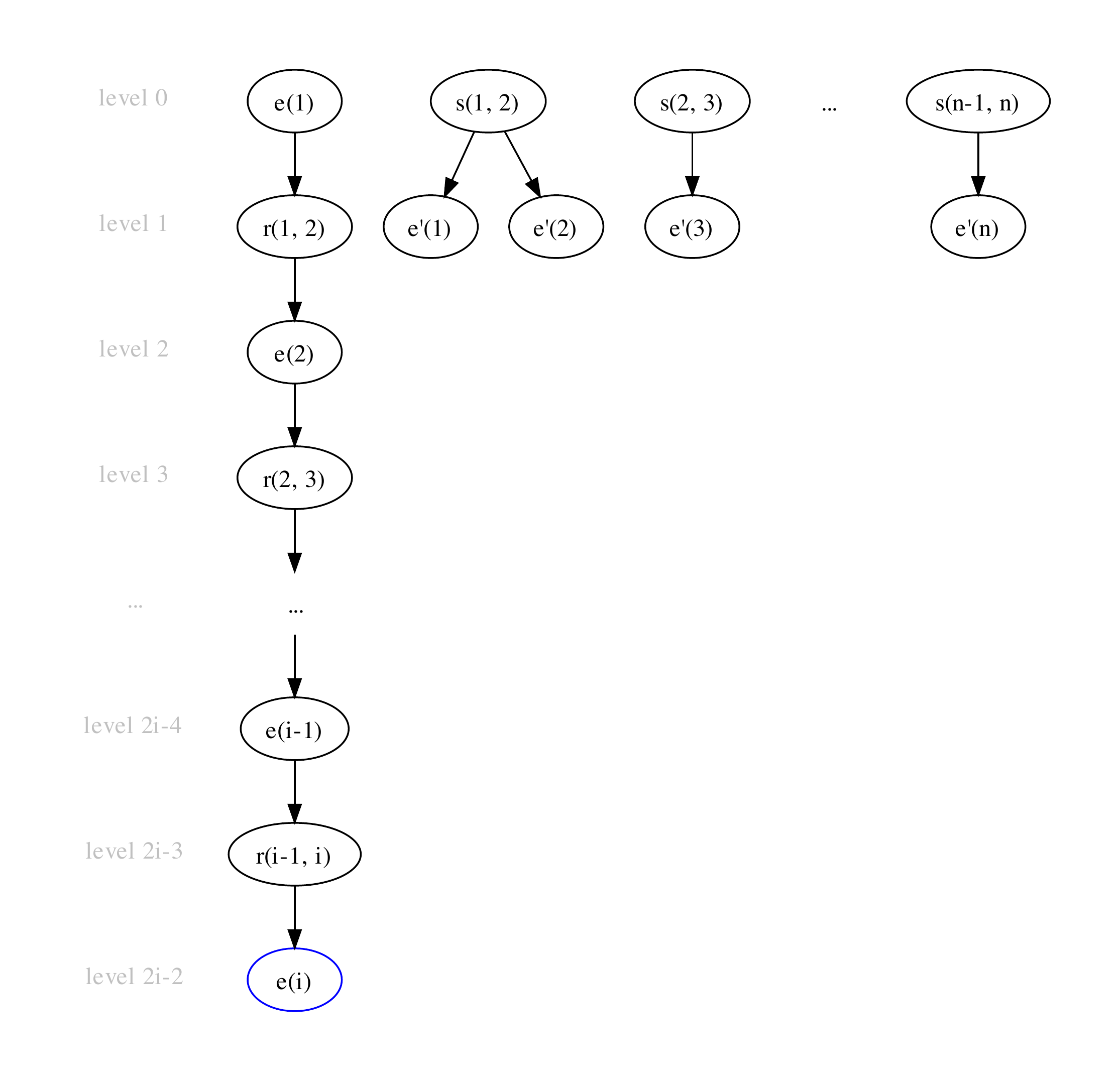}
       \caption{$e(i)$ is generated at level $2i-2$}\label{fig:chase4}
    \end{subfigure}
\end{figure}

\section{Discussion on related and future work}

Incompleteness in databases is a central topic in the field of logic in databases~\cite{CLM:JAL2010,CLM:SIGREC2009}.
Data incompleteness is likely to occur in several application scenarios, such as data integration.
When querying incomplete data, reasoning on the schema is often necessary in order to provide the correct answers.
A query answering algorithm addressing incomplete data under constraints is described in~\cite{CM:TPLP2010}. There, the schema is expressed with an extended version of the Entity-Relationship model, and the initial query is rewritten as a recursive Datalog query that encodes the information about the schema.
The extension of the Entity-Relationship (ER) model~\cite{Chen76} used here is called Extended ER (EER) model~\cite{Cali06}.
This model is also an extension of the model presented in~\cite{CCDL01e}.
Here, we focused on some aspects related to the techniques for answering queries under the dependencies, called Conceptual Dependencies, enforced by the EER model.
Considering the certain answers under the so-called \emph{sound semantics} (see, e.g., \cite{CaLR03}) requires proper attention in order to answer queries correctly~\cite{ArBC99}.
In particular, query answering under dependencies makes use of the notion of \emph{chase}~\cite{MaMS79,MaSY81,Vard83,JoKl84}.

Johnson and Klug~\cite{JoKl84} proved that, in order to test containment of CQs under IDs alone or \emph{key-based} dependencies (a special class of KDs and IDs), it is sufficient to consider a \emph{finite}, initial portion of the chase.
The result of~\cite{JoKl84} was extended in~\cite{CaLR03} to a broader class of dependencies, strictly more general than keys with foreign keys: the class of KDs and \emph{non-key-conflicting inclusion dependencies (NKCIDs)}~\cite{Cali03t}, that behave like IDs alone because NKCIDs do not interfere with KDs in the construction of the
chase.
The above results about query containment (see, e.g.,~\cite{CaGK08}) can be straightforwardly adapted to solve the decision problem of answering on incomplete databases, since the chase is a representative of all databases that satisfy the dependencies and are a superset of the initial data.

In a set of CDs, IDs are not non-key-conflicting (or better \emph{key-conflicting}), therefore the decidability of query answering cannot be deduced from~\cite{JoKl84,CaLR03}, (though it can be derived from~\cite{CaDL98}).  In particular, under CDs, the construction of the chase has to face interactions between KDs and IDs.
In spite of the potentially harmful interaction between IDs and KDs, analogously to the case of IDs alone~\cite{CCDL04}, in the presence of CDs, the chase is a representative of all databases that are a superset of the initial (incomplete) data, and satisfy the dependencies.

Future work includes the extension of the applicability of chase-based techniques to further classes of constraints.

Relevant directions of research regard all those area in which integrity constraints are used to characterize useful scenarios in which query answering plays an important role.
Among these, we mention \emph{access patterns}, which are constraints indicating which attributes of a relation schema are used as input and which ones are used as output.
In this respect, access patterns may suitably characterize several relevant contexts, such as Web forms, legacy data, Web services, and the so-called Deep Web~\cite{CM:APWEB2010,CM:EDBT2010,M:CRYPT2011}. Query processing under access patterns requires specialized techniques.
Among these, static optimization, including query containment via techniques similar to those described in this paper, has been studied for limited forms of conjunctive queries~\cite{CM:ER2008,CM:ICDE2008,CCM:EROW2007,CMC:SEBD2007}. More general cases are covered in the context of dynamic optimization~\cite{CCM:JUCS2009}, where results are available for schemata with functional dependencies and simple full-width inclusion dependencies. The latter kind of dependencies, albeit simple, can be used to state equivalence, and thus captures the notion of relations with multiple access patterns.

Another context where integrity constraints play a major role is the orthogonal dimension of integrity constraint checking. In the context of relational as well as deductive databases, correct and efficient integrity checking is a crucial issue: without any guarantee of data consistency, the answers to queries cannot be trusted.
Checking integrity constraints from scratch may be prohibitively time consuming, as databases may contain huge quantities of data.
However, a procedure that generates ``simplified'' incremental checks for given update patterns can be adopted: simplified versions of the constraints can be automatically derived at database design time and tested before the execution of any update. In this way, virtually no time is spent for optimization or rollbacks at run time~\cite{CM:FI2006,DM:IGP2008,MCD:IGP2006,CM:LPAR2005,MC:ADBIS2005,MC:DEXA2005,M:ADBIS2004,M:FQAS2004,CM:LOPSTR2003,M:PHD2005,martinenghi2003simplification,M:CORR2013a}.
The simplification procedure may also be adapted to several other contexts, such as data integration systems~\cite{CM:FoIKS2004}, automatic generation of repairs for inconsistent data~\cite{CM:LAAIC2006}.
It is also possible to reconsider the whole approach in an ``inconsistency-tolerant'' way, 
i.e., without requiring full data integrity (which is indeed very unlikely in real cases): in this case one 
can guarantee, through simplified checking, that no new inconsistencies are introduced by updates~\cite{DM:TKDE2011,DM:PPDP2008,DM:LPAR2006,DM:ADVIS2006,DM:SEBD2006,DM:QOIS2009,DM:FlexDBIST2007,DM:TDM2006,DM:FlexDBIST2006,DM:IGP2009}.

Other kinds of constraints may occur at the query level, for instance when the constraint specifies a limit on the number of results that the query should return, although many more satisfy the query. It should be interesting to see whether there is any relationship whatsoever between the constraints in the logical sense described here and the constraints on the query results of these other works.
When posing a query over multiple sources, a user is often interested in determining the $k$ most relevant results that match given conditions. Relevance is usually expressed as 
a function that combines the scores of the data from each single source into an aggregate score.
The naive approach to address these queries consists in first computing \emph{all} the query results, then sorting them by relevance.
This process is very expensive.
Fortunately, the sources are often endowed with special access modes that allow retrieving only a small fraction of the available tuples, yet guaranteeing that the top $k$ results are found.
Investigations on top-$k$ query scenarios have abounded in the recent years.
In \emph{proximity rank join}~\cite{MT:TODS2012,MT:PVLDB2010}, the objects returned by the sources are equipped with a score as well as with a real-valued feature vector, which represents the ``geometry'' of the problem, e.g., the location of the object in the space. Here, the vector space plays a distinctive role in the computation of the overall score of a result and makes the problem more challenging than in the traditional case. 
In the same setting, one may additionally wish to diversify the result set, yet retaining only results with high scores~\cite{CCFMT:TODS2013,FMT:SIGMOD2012,DBLP:conf/sebd/FraternaliMT12}.
When multiple sources are joined, and both random and sorted accesses are available, suitable execution strategies can be devised so as to further speed up the computation of the top $k$ results~\cite{MT:TKDE2011,CMT:TECHREP2009}.
The topology of the join between two sources (in parallel or in a sequence) is also a relevant factor that determines the most promising execution strategy for a top-$k$ query~\cite{MT:DBRANK2010}.
Often, users are unable to precisely specify the scoring functions (e.g., weighted sums) used to rank the results of a query.
Adopting uncertain/incomplete scoring functions (e.g., weight ranges) can better capture user's preferences.
Semantics of ranking queries and sensitivity of computed results to refinements made by the user in the presence of uncertainty are studied in~\cite{SIMT:SIGMOD2011}.
All these optimization opportunities are especially relevant in the context of search~\cite{IMT:SECO2009}

Yet another kind of constraint that is used to complete the semantics of a query by means of a sort of query ``expansion'' is given by taxonomies and ontologies. Traditional information search, in which queries are posed against a known and rigid schema over a structured database, is shifting towards a Web scenario in which exposed schemas are vague or absent, and data comes from heterogeneous sources. In this framework, query answering cannot be precise and needs to be relaxed, with the goal of matching user requests with accessible data.
Suitable models and languages are needed for querying data sets with vague schemas.
When additional information about the data is available (in the form of simple classifications of terms arranged in a hierarchical structure or contextual information), extensions of relational algebra addressing these issues become possible~\cite{MT:ER2010,MT:FQAS2009,MT:ITAIS2010,MT:ITAIS2009,MT:TECHREP2009}.
Taxonomical information can also be provided via the notion of context. When answering a query, it is important to remove all the data that are not relevant with respect to the context in which they are used. This process, known as context-aware data tailoring, is obtained in~\cite{RMT:LID2011} via Answer Set Programming techniques.

Constraints may also occur in logic programming, where constraint programming techniques are use to enable meta-programming paradigms endowed with features such as reversibility of a meta-interpreter, which turns it into a powerful program generator, as well as incremental evaluation of integrity constraints~\cite{CM:AAI2000}.







\bibliographystyle{abbrv}
\bibliography{ChaseDataDependency}







\end{document}